\documentclass[twocolumn,showpacs,pra,aps]{revtex4-1}
\usepackage{graphicx}
\usepackage{epstopdf}
\usepackage{amsmath}
\usepackage{amssymb}
\usepackage{epsfig}
\usepackage{amsthm}
\usepackage{color}
\usepackage{textcomp}
\begin{document}

\title{Single-photon stimulated four wave mixing at telecom band }

\author{Shuai Dong$^{1,\dagger}$ }
\author{Xin Yao$^1$}
\author{Wei Zhang$^{1,*}$}
\author{Sijing Chen$^2$}
\author{Weijun Zhang$^2$}
\author{Lixing You$^2$}
\author{Zhen Wang$^2$}
\author{Yidong Huang$^1$}

\affiliation{$^1$Tsinghua National Laboratory for Information Science and Technology, Department of Electronic Engineering, Tsinghua University, Beijing, 100084, China}
\affiliation{$^2$State Key Laboratory of Functional Materials for Informatics, Shanghai Institute of Microsystem and Information Technology, Chinese Academy of Sciences, Shanghai 200050, China}
\affiliation{$^\dagger$dongshuai2008@gmail.com}
\affiliation{$^*$Corresponding author: zwei@tsinghua.edu.cn}

\begin{abstract}
Single-photon stimulated four wave mixing (StFWM) processes have great potential for photonic quantum information processing, compatible with optical communication technologies and integrated optoelectronics. In this paper, we demonstrate single-photon StFWM process in a piece of optical fiber, with seeded photons generated by spontaneous four wave mixing process (SpFWM).  The effect of the single-photon StFWM is confirmed by time-resolved four-photon coincidence measurement and variation of four-photon coincidence counts under different seed-pump delays.  According to the experiment results, the potential performance of quantum cloning machine based on the process is analyzed. 
\end{abstract}

\maketitle

Nonlinear optical parametric processes, including parametric down conversion and four wave mixing, have been playing important roles in researches of quantum optics and in the development of photonic quantum information processing. Two types of parametric processes are concerned in these applications, spontaneous one and stimulated one, depending on different initial conditions. In spontaneous parametric processes, only pump light is injected into nonlinear media, and signal and idler photons are generated from the vacuum fluctuation with quantum correlation/entanglement. In stimulated parametric processes, not only the pump light but also seeded photons are injected into nonlinear media, and new pairs of photons would be generated, correlated with the seeded photons. Both the two types can be used for quantum information processing, for example, the spontaneous parametric down conversion (SPDC) has been widely used to generate various photonic quantum states \cite{polEntSPDC,franson1989bell,timebin,fourPhotonEntanglement,frequencyBin,fibPolEnt,hyperEntSPDC} , providing an important way to realize photonic quantum states, and the stimulated parametric down conversion has been utilized to realize optimal quantum optical cloning \cite{quantumCloning2} and quantum NOT gate \cite{quantumNotGate} and to generate multi-photon entangled states \cite{ampQuanEntg}. 

In recent years, the applications of four wave mixing processes in photonic quantum information processing have attracted much attention, since they can be realized in third-order waveguides such as optical fibers and silicon wire waveguides, which can be operated at telecom band and are well compatible with developed technologies of optical communications and integrated optoelectronics \cite{fibPolEnt,energyTimeSpFWM,nanowireSpFWM,microringSpFWM,chipSpFWM,chipSpFWM2}. In the four wave mixing process, pairs of pump photons are annihilated, and pairs of correlated photons are generated, which are usually named as signal and idler photons, respectively. In interaction picture, the simplified Hamiltonian for four wave mixing in an isotropic media can be expressed as
\begin{equation}
\hat{H}_{int}\sim k\hat{a}^\dagger_s\hat{a}^\dagger_i+h.c.,
\label{eq:FWM-H}
\end{equation}
where $k$ is a nonlinear gain parameter, proportional to the nonlinear parameter $\chi^{(3)}$, and also include the effect of phase matching condition, $\hat{a}_s^\dagger$($\hat{a}_i^\dagger$) is the generation operator of signal(idler) photons, h.c. stands for Hermitian conjugate. Undepleted pump approximation is applied in the derivation of Eq. \ref{eq:FWM-H} \cite{SpFWM_fiber_1}.

In spontaneous four wave mixing (SpFWM) processes, correlated photon pairs are generated with vacuum state as the initial state, i.e. 
\begin{equation}
\vert \Psi_1\rangle=\exp(-i\hat{H}_{int}t)\vert 00\rangle=\vert 00\rangle+g\vert11\rangle,
\label{eq:state-SpFWM}
\end{equation}
where $\vert mn\rangle$ is photonic quantum state with $m$ photons in the signal mode, and $n$ photons in the idler mode. $g=ikt$, where $t$ is the interaction time.  Higher order terms have been neglected for $g\ll1$. Based on this process, various complex photonic quantum states have been realized at telecom band by coherent manipulation and superposition of the generated correlated state, such as polarization entangled state \cite{fibPolEnt,fibPolEnt2}, frequency entangled state \cite{freqEntSpFWM} and hyper-entangled state \cite{hypEntSpFWM}.
In stimulated four wave mixing (StFWM) processes, the initial states of the signal and idler modes are no longer vacuum states.  In this paper, we focus on the StFWM process with single photons as the initial state. Assuming that there are single photons in the signal mode as the seeded photons, i.e., the input state is $\vert 10\rangle$. By the four wave mixing process, the output state can be expressed as
\begin{equation}
\vert \Psi_2\rangle=\exp(-i\hat{H}_{int}t)\vert 10\rangle=\vert 10\rangle+\sqrt{2}g\vert21\rangle.
\label{eq:staet-StFWM}
\end{equation}
Comparing with the output state of the SpFWM process shown in Eq.\ref{eq:state-SpFWM}, there is an increase in the probability of generating photon pairs by a factor of 2 due to the stimulating effect of the seeded photons. The stimulated signal photons are in the same mode with the seeded signal photons, and the idler photons are also correlated with the two signal photons. Hence, it can be expected that the StFWM process has the potential for photonic quantum state cloning and multi-photon correlated/entangled states generation. 
However, the realization of single photon StFWM is not easy. The third-order nonlinearity is far weaker than the second-order one. On the other hand, usually the frequency difference between the signal/idler photons and the pump light is small in four wave mixing processes, which leads to difficulties in noise photon suppression. Hence, there is no experiment demonstration of the single-photon StFWM in third-order waveguides as far as the authors’s knowledge.

In this paper, we demonstrate the single-photon StFWM process in a piece of optical fiber. The noise photons generated by spontaneous Raman scattering are highly suppressed by cooling the fiber to a temperature of 2 K \cite{SpFWM_CoolingFiber,SpFWM_CoolingFiber2}. The seeded single photons are signal photons from correlated photon pairs generated via SpFWM in the same piece of optical fiber. After the StFWM process, new pairs of photons are generated and the four photons are correlated. By carrying out time-resolved four-photon coincidence measurement, all the coincident and accidental coincident counts can be obtained and analyzed in the experiment. The effect of single-photon StFWM is confirmed by the coincidence to accidental coincidence ratio and by the variation of the four photon coincidence under different seed-pump time delay. According to the experiment results of the single-photon StFWM process, the potential performance of quantum cloning is analyzed. 
 
%\section{Experiment and results}
\begin{figure}[htbp]
\centering
\includegraphics[width=\linewidth]{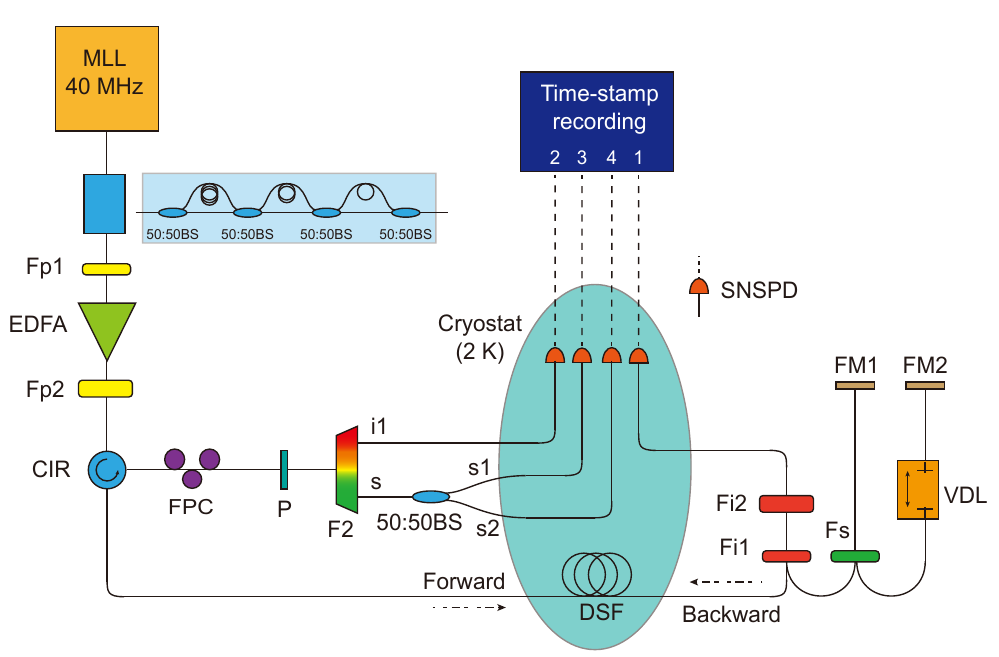}
\caption{The experiment setup. MLL: mode-locked pulse laser; 50:50BS: 50\%:50\% optical beam splitter; Fp1, Fp2, Fi1, Fi2, Fs, Fsi , F: filters realized by (cascaded) 100GHz DWDMs; CIR: optical circulator; FPC: fiber polarization controller; FM: Faraday mirror; VDL: variable optical delay line; Pol.:polarizer; SNSPD: super-conducting nanowire single photon detector.}
\label{fig:setup}
\end{figure}

The experiment setup is shown in Fig \ref{fig:setup}. The pump source is realized by a mode-locked fiber laser, which has a repetition rate of $40$ MHz, and three unbalanced Mach-Zenhder interferometers (MZIs) made by 50:50 optical beam splitters (50:50BS). The three unbalanced MZIs have different time delays between their arms, which is about 12.5ns, 6.2ns and 3.1ns, respectively. They are cascaded to multiplex the light pulses, leading to a repetition rate of 320MHz, aimed at high photon generation rates.  A 100 GHz dense wavelength division multiplexer (DWDM, Fp1) is used as an optical filter to select the pump light with center wavelength ($1552.52$ nm) and narrow the spectrum ($0.3$ nm). After amplified by an Erbium doped fiber amplifier (EDFA), a high extinction ratio (>120 dB) optical filter (Fp2) made by cascaded DWDM devices is used to suppress noise photons at the signal and idler wavelengths. The pump pulses are injected into a piece of dispersion shifted fiber (DSF) through an optical circulator (CIR) to generate correlated photon pairs via the SpFWM process. The DSF, with a length of 280 m, is placed in a cryostat for superconducting nanowire single photon detectors (SNSPDs) and cooled to 2 K. Such a low operation temperature is helpful to suppress noise photons generated by the spontaneous Raman scattering in the fiber effectively \cite{SpFWM_CoolingFiber,SpFWM_CoolingFiber2}. The filters for signal photons (Fs, central wavelength: 1549.32 nm) and for idler photons (Fi1, central wavelength 1555.75 nm) are used to separate the signal photons, idler photons and residual pumps. The idler photons are detected using a SNSPD after a high-extinction-ratio filterer (Fi2). The related filters are made by DWDM devices. The signal photons and the residual pump pulses are reflected back to the DSF by Faraday mirrors (FM1, FM2), which also guarantee that their polarization states are aligned when they propagate backward in the DSF \cite{FM_Pol}. Through a variable optical delay line (VDL), the pump pulses have an adjustable time delay relative to the reflected signal photons. The reflected signal photons, as seeded photons, would stimulate the four wave mixing process backward pumped by the reflected pumps when they are temporally overlapped.

The seeded photons, the stimulated signal and idler photons are separated by a filter module (F2) for signal and idler wavelengths. Before the filter module, a polarizer (P) and a fiber polarization controller (FPC)are used to block the perpendicular-polarized Raman noise photons due to the fiber pigtails of optical devices outside the cryostat. A 50:50 BS is used to split the seeded and the stimulated signal photons and direct them to two SNSPDs probabilistically. The total detecting efficiency is about $10\%$ for each channel, including the loss of filters and FPCs, detecting efficiency of SNSPDs, after optimized by adjusting the polarization states of photons in each channel using FPCs (not shown in Fig. \ref {fig:setup} for simplicity). The photon counting rate in each channel is about $1$ MHz.  After detected by SNSPDs, their channel number and arrival time-stamps are recorded using a four-channel time correlated single photon counting module (TCSPC). The data are post-processed using a personal computer to fulfill the time-resolved four-photon coincidence measurement.

\begin{figure}[htbp]
\centering
\includegraphics[width=\linewidth]{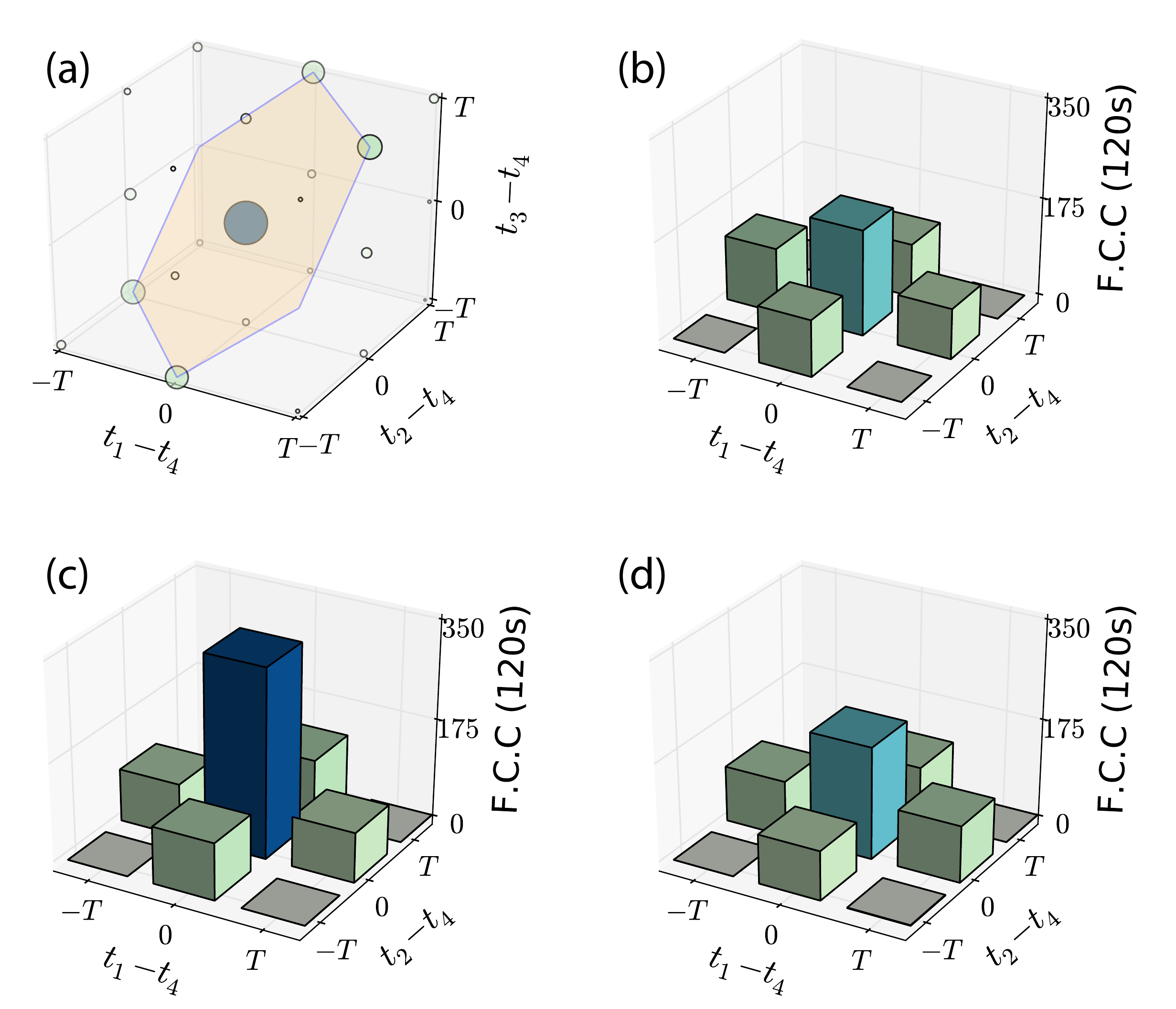}
\caption{Time-resolved four-photon coincidence measurement results. (a), four-photon (accidental) coincidence distribution related to the time-differences of the recorded single-photon events ($t_i$: time stamps of single-photon events in the Channel $i$). Only at some specific bins there will be an obvious four-photon (accidental) coincidence counts. (b)(c)(d), four photon (accidental) coincidence distribution in the plane $t_3=t_1+t_2-t_4$. (b),(c) and (d) are related to different seed-pump delays $\tau$. (b) $\tau$=-18.98 ps; (c) $\tau$=0.62 ps; (d) $\tau$=21.62 ps.}
\label{fig:fpfcc}
\end{figure}

Since there are four channels, three time-differences, for example, $t_1-t_4$, $t_2-t_4$, and $t_3-t_4$, are used as temporal variables to depict the four-photon coincidence completely. Here, $t_i (i=1, 2, 3 {\textrm{ or }} 4)$ is the recorded time stamps of photons in Channel $i$ as labeled in Fig.\ref{fig:setup}. By post-processing the time-stamps of photons in the four channels, we can calculate the four-photon coincidence counts in any range of $t_1-t_4$, $t_2-t_4$, and $t_3-t_4$. In the experiment, the time-stamps are recorded with a resolution of $1$ ps, while the timing jitters of the SNSPDs are about $70$ ps. In the calculation of the coincidence counts, the time bin width is set to $2.56$ ns. The time-resolved four-photon coincidence measurement results are shown in Fig. \ref{fig:fpfcc}. Figure \ref{fig:fpfcc}(a) is a typical four-photon coincidence distribution related to the three time-differences among photons in the four channels. The bubble size is proportional to the four-photon (accidental) coincident counts. It is obvious that the four-photon (accidental) coincident counts are much larger than zero only at some specific bins. These bins satisfy the relation,
\begin{eqnarray}
(t_1-t_4)+(t_2-t_3)&\equiv&(t_1-t_3)+(t_2-t_4)=0,\\ \nonumber &&t_i=nT, i\in\{1,2,3,4\},n\in \mathbb{Z},
\label{equ:fccPlane}
\end{eqnarray}
where $T$ is the repetition period of the pump pulses. Eq. \ref{equ:fccPlane} indicate that the four-photon accidental coincidence counts are mainly due to pairs of correlated photons generated from different pump pulses by SpFWM or StFWM. The contributions of the noise photons and dark counts of the SNSPDs are quite small and can be neglected. Hence, we only focused on the coincidence counts in these specific time bins satisfying Eq. \ref{equ:fccPlane} which is a scalar equation of a plane in Fig. \ref{fig:fpfcc}(a) . We plot the distributions of four-photon coincidence in these time bins in Fig. \ref{fig:fpfcc}(b, c, d), in which the different time delays between the seeded signal photons and the reflected pump pulses are set. This time delay is denoted by $\tau$ and can be adjusted by the VDL.  Fig. \ref{fig:fpfcc}(b), (c) and (d) shows the experiment results in the cases that the seed-pump time delays  are $\tau$=-18.98 ps, $\tau$=0.62 ps  and $\tau$=21.62 ps, respectively. For Fig. \ref{fig:fpfcc}(b) and (d),  the time delays are larger than the coherence time of the seeded signal photons and the pump pulses, which are determined by the bandwidths of corresponding optical filters. The central bars represent the four photon coincident counts in the time bin of $t_1-t_4=t_2-t_4=t_3-t_4=0$. It is defined as the coincident count (CC), since in this case the four photons are generated by the same pump pulse, i.e. a pair in the forward process and another pair in the backward process. In the figures, coincident counts in the time bins $(\pm T,0)$,$(0,\pm T)$ are also much larger than zero, where $(x,y)$ represents the bins at $t_1-t_4=x,t_2-t_4=y$ (and $t_3-t_4=x+y$). Their average counts of these bins are defined as the accidental coincident count (ACC), since in these cases the four photon coincidence counts are due to pairs of photons generated from different pump pulse. The ratio of CC and ACC is defined by
\begin{eqnarray}
R=\frac{1}{2}\frac{CC}{ACC},
\end{eqnarray}
where the factor $\frac{1}{2}$ is due to the probabilistic properties of splitting two signal photons at the 50:50 BS. According to the experiment data, the ratios are $R=0.94$ for $\tau=-18.98$ ps, and $R=1.04 $ for $\tau=21.62$ ps, agreeing well with the theoretical value $R_{\overline{\textrm{StFWM}}}=1$ for the case that there is no StFWM and the photon pair generation backward in the DSF is due to the SpFWM  (Shown in the Supplementary Material).

The condition is different for Fig. \ref{fig:fpfcc} (c), in which the seed-pump delay is $\tau=0.62$ ps. In this case the reflected seeded photons and pump pulses are overlapped temporally, there will be an increase in the probability of generating new pairs of photons in the backward process, due to StFWM process. Theoretical analysis (Shown in the Supplementary Material) shows that the CC to ACC ratio should be $R_{\textrm{StFWM}} = 2$ if the StFWM occurs in the backward process. The measured value is $R =1.71$ according to the experimental results in Fig. \ref{fig:fpfcc}(c), confirming the existence of the single-photon StFWM process. The deviation between the experimental results and the theoretical values is mainly due to the imperfect match between the seeded photons and the pump pulses in time domain and polarization.

\begin{figure}[htbp]
\centering
\includegraphics[width=\linewidth]{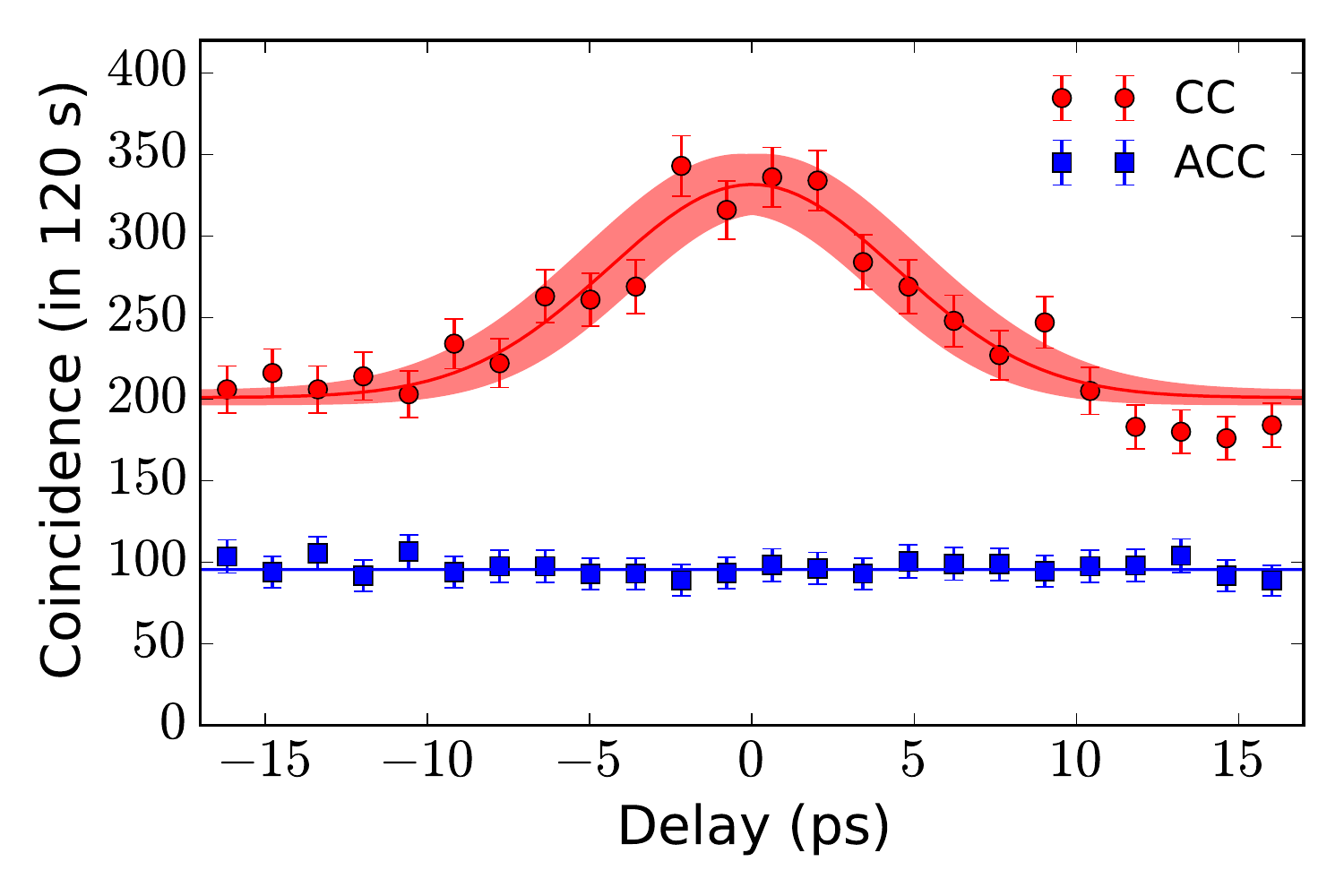}
\caption{Four-photon coincidence counts (red) and four-photon accidental coincidence counts (blue) vary with different seed-pump delays.}
\label{fig:delayFcc}
\end{figure}

We also measured the CC and ACC under different seed-pump delay $\tau$ by adjusting the VDL. The results are shown in Fig. \ref{fig:delayFcc}, in which the red circles and blue squares denote the CCs and ACCs, respectively. The red line is the fitting curve of CCs, which has a Gaussian shape, the blue horizontal line is the fitting line of the ACCs. The red shadow around the fitted curve of the CCs is the region covering the fitting curves with all the fitting parameters varying in a $\pm \sigma$ interval, where $\sigma$ stands for their standard variances. It can be seen that when the seed-pump delay approaches zero, the CC increases obviously, and achieves its maximum at $\tau=0$. The obvious enhancement of the CC only occurs when the time delay $\tau$ is in a range of about $10$ ps, which agrees well with the coherence times of the seeded photons and the pump pulses. On the other hand, the ACCs are independent on the seed-pump time delay with small fluctuations only due to the experimental errors.

According to the experimental results shown in Fig. \ref{fig:delayFcc}, an important parameter which is related to the performance of quantum cloning machine based on the StFWM process can be calculated as following \cite{quantumCloning2},

\begin{equation}
R'=\frac{C_{\tau=0}}{C_{\tau\gg t_c}}=1.66,
\end{equation}
where $C_{\tau=0}$ is the CC at $\tau=0$, $C_{\tau\gg t_c}$ is the CC when $\tau$ is much larger than the coherence time of seeded photons and pump pulses $ t_c $. It can be seen that the value agrees well with the CC to ACC ratio $R$. Both $R$ and $R'$ can be used to confirm the existence of StFWM. It should be noted that the CC to ACC ratio $R$ can be calculated only using the experiment results when $\tau$ is close to zero, hence it can be taken by one-time four-photon coincident counting measurement, without tuning the time delay $\tau$. Hence, the time-resolved four-photon coincidence measurement is helpful to simply the experiment, by which the seeded photons and the pump pulses don’t need to be split and reflected separately, and the VDL is not required.

According to the value of $R'$, the cloning fidelity $F$ can be estimated if the experiment setup in this paper is applied to quantum cloning \cite{quantumCloning2},
\begin{equation}
F=\frac{2R'+1}{2R'+2}=81.2\%,
\end{equation}
which is very close to the theoretical upper bound of cloning of one qubit to obtain two qubits,  $83.3\%$ \cite{quantumCloning2}. It can be seen that the single-photon StFWM process has great potential for quantum information processing based on quantum cloning, especially the applications at telecom band.

%\section{Conclusion}
As a summary, in this paper we have demonstrate StFWM process in nonlinear dispersion shifted fibers, which is placed in a 2 K cryostat to suppress the Raman noise effectively. Single photons from correlated photon pairs generated by SpFWM process is used as the seeded photons in StFWM process. We characterize the StFWM  process by the time-resolved four-photon coincidence measurement, in which four-photon coincidence and four-photon accidental coincidence can be calculated. The StFWM process is demonstrated by $R = 1.71 > 1$ when the seeded photons and pump pulses are overlapped and by the variation of four-photon coincident counts under different seed-pump delays. There is an increase by a factor of 1.66 in the four-photon coincidence count when the delay is $\tau=0$ compared with the ones with $\vert \tau\vert\gg $ coherence time of the generated photons and the pump pulses. The potential performance of quantum cloning machine based on the StFWM process has also been analyzed and the cloning fidelity for one qubit to two qubits case is estimated to be about $81.2\%$ according to the increase factor of $1.66$, which is close to the theoretical upper bound. 

\section*{Funding Information}
This work was supported by 973 Programs of China under Contract No. 2013CB328700 and 2011CBA00303, the National Natural Science Foundation of China under Contract No. 61575102, 91121022 and 61321004, Tsinghua University Initiative Scientific Research Program under Contract No. 20131089382 and Strategic Priority Research Program (B) of the Chinese Academy of Sciences (XDB04020100).
%\section*{Acknowledgments}

\clearpage
\section*{Supplimental: The analysis of the CC to ACC ratio}
The four-photon coincident count (CC) is resulted from two photon pairs generated by a single pump pulse, one pair by SpFWM process in the forward direction, and one pair by SpFWM or StFWM in the backward direction, depending on the time delay between the seeded photons and the pump pulses. Assuming that $R_1$ is the photon pair generation rate in the forward direction, and $R_2$ is that in the backward process, when the time delay between the seeded photons and the pump pulses is much larger than their coherence time, i.e., both the forward and the backward processes are the SpFWM processes. The four-photon CC rate will be
\begin{eqnarray}
R_{CC}=\frac{1}{2}\eta R_1 R_2.
\end{eqnarray}
The denominator 2 comes from the fact that there is half possibility that the two signal photons will go to the same output port of the 50:50 BS and no four-photon coincidence count can be expected in this case.$\eta$ is a coefficient related to the system collection efficiency, including the channel loss and detecting efficiencies of the SNSPDs. 

While the accidental coincident counts (ACCs) are due to two pairs of photons generated by different pump pulses, one in the forward direction and another in the backward direction. The ACCs shown in Fig. 2 (in the main text) come from that two pairs of photons are generated by two subsequent pump pulses.  Assuming that there are two subsequent pump pulses, indicated by pulse 1 and pulse 2 with pulse 1 in advance of pulse 2 and that both pulses generate a pair of photons by SpFWM process. The signal and idler photons generated by the pump pulse 1(2) are indexed by $s_1$($s_2$) and $i_1$($i_2$), respectively.  The channels through which photons will go out and related probabilities are shown in Table S1. The cases that four photons detected in four different channels would result in four-photon ACC at specific time bins in Fig.2, which are indicated by  $(x,y),x, y\in \{0,\pm T\} $ with $t_1-t_4=x,t_2-t_4=y$ (and $t3-t4=x+y$).
\begin{table*}
\centering
\caption{{\bf All the possible outputs of the four photons and corresponding possibilities and the bins of accidental coincident counts.}}
\begin{tabular}{|l*{8}{|c}|}
\hline
&\multicolumn{4}{c|}{Forward $s_1$\&$i_1$, backward $s_2$\&$i_2$}& \multicolumn{4}{c|}{Backward $s_1$\&$i_1$, forward $s_2$\&$i_2$}\\
\hline
$i_1$ &1 &1 &1	&1 &2 &2 &2	&2\\
\hline
$s_1$ &3 &3 &4	&4 &3 &3 &4	&4\\
\hline
$i_2$ &2 &2 &2	&2 &1 &1 &1	&1\\
\hline
$s_2$ &4 &3 &3	&4 &4 &3 &3	&4\\
\hline
Probability	&1/4 &1/4 &1/4 &1/4 &1/4 &1/4 &1/4 &1/4\\
\hline
ACC bin &$(-T,0)$ &No ACC &$(0, T)$ &No ACC &$(0, -T)$ &No ACC &$(T,0)$ &No ACC\\
\hline
\end{tabular}
  \label{table}
\end{table*}

The four terms of ACC in Table \ref{table} is related to the four-accidental-count bars in Fig. 2(b,c,d). Hence, accoding to Table \ref{table}, for every ACC bar, the four-photon ACC rate is 
\begin{eqnarray}
R_{ACC}=\eta R_1 R_2/4.
\end{eqnarray}

According to the definition of CAR (Eq.(5)), for the case when only SpFWM process occurs in the backward direction, the theoretical prediction of CAR would be 
\begin{eqnarray}
R_{\overline{\textrm{StFWM}}}=\frac{1}{2}\frac{R_{CC}}{R_{ACC}}\equiv 1.
\end{eqnarray}

On the other hand,if the seeded photons and the pump pulses are overlapped temporally, the StFWM process would occur in the backward direction, hence, the generation rate of photon pairs in the backward direction would increase by a factor of 2, i.e., $R_2\to R_2^*=2R_2$, according to the Eq. (3). While, the ACC rates remain unchanged. Hence, the theoretical prediction of CAR, when StFWM process occurs in the backward direction, would be
\begin{eqnarray}
R_{\textrm{StFWM}} =\frac{1}{2}\frac{\eta R_1 R_2^*/2}{\eta R_1 R_2/4}\equiv 2.
\end{eqnarray}

\end{document}